\documentclass[12pt,epsfig]{article}
\usepackage{graphicx,amsmath,amssymb}

\newlength{\dinwidth}
\newlength{\dinmargin}
\def\lapproxeq{\lower .7ex\hbox{$\;\stackrel{\textstyle
<}{\sim}\;$}}
\def\gapproxeq{\lower .7ex\hbox{$\;\stackrel{\textstyle
>}{\sim}\;$}}
\def\gtrsim{\lower .7ex\hbox{$\;\stackrel{\textstyle
>}{\sim}\;$}}
\def\lesim{\lower .7ex\hbox{$\;\stackrel{\textstyle
<}{\sim}\;$}}

\def\be{\begin{equation}}
\def\ee{\end{equation}}
\def\bea{\begin{eqnarray}}
\def\eea{\end{eqnarray}}

\def\ra{ \rightarrow }

\begin{document}
\begin{flushright}
IPPP/06/29 \\
DCPT/06/58 \\
7th August 2006 \\

\end{flushright}

\vspace*{0.5cm}

\begin{center}
{\Large \bf Central jet production as a probe of the perturbative formalism
  for exclusive diffraction}

\vspace*{1cm}
\textsc{V.A.~Khoze$^{a,b}$, A.D. Martin$^a$ and M.G. Ryskin$^{a,b}$} \\

\vspace*{0.5cm}
$^a$ Department of Physics and Institute for
Particle Physics Phenomenology, \\
University of Durham, DH1 3LE, UK \\
$^b$ Petersburg Nuclear Physics Institute, Gatchina,
St.~Petersburg, 188300, Russia \\

\end{center}

\vspace*{0.5cm}

\begin{abstract}

We propose a new variable, $R_j$, in order to
identify exclusive double-diffractive high $E_T$ dijet production. The variable $R_j$ is
calculated using the transverse energy
$E_T$ and pseudorapidity of the jet with the {\it largest} $E_T$. 
For a purely exclusive event the value of $R_j\to 1$, if we were to neglect hadronisaton and the
detector resolution effects. To illustrate
the expected $R_j$-distribution we also compute exclusive three-jet
production; and, moreover, include jet smearing effects. 
By studying the predictions as a function of the size of the rapidity interval, $\delta\eta$,
which allows for additional gluon radiation, one can probe the QCD
radiation effects which are responsible for the Sudakov suppression
of the exclusive amplitude. In this way we may check,
and improve, the formalism used to predict the cross sections of
exclusive double-diffractive Higgs boson (and/or
other New Physics) production.

\end{abstract}

\section{Introduction}
Diffractive processes offer a unique means to discover new
physics at the LHC, see for example, \cite{ar,KMRProsp,DKMOR,cox}. An
exciting possibility is to search for Higgs bosons in an exclusive reaction, that is
$pp \to p+H+p$, where the plus signs denote large rapidity gaps.
This process allows detailed measurements of the Higgs boson properties in an exceptionally
clean environment and provides a unique signature, especially for the MSSM Higgs sector,
see \cite{KKMRext, Georg}. In particular, the Higgs mass and spin-parity determination
can be done irrespective of the decay mode, and these studies are at the heart
of the recent proposal \cite{LOI} to complement the central detectors at the LHC
by forward proton taggers placed far away from the interaction point.
However, the expected event rate is limited;
 it is strongly suppressed, in particular by a Sudakov form factor necessary to guarantee the exclusive final state, see for instance \cite{KMR,jeff}.
An analogous Sudakov suppression enters the predictions for the exclusive production of dijets, $\gamma\gamma$, etc.
The existing diffractive Tevatron data (see, for example, the reviews \cite{KG1,KG2,MG1,CM,MG2,koji}
and references therein) are not in disagreement with the theoretical expectations
for these processes, see \cite {KKMR,KKMRdj,KMRS,MKKRS, BH75}. However a
definitive\footnote{The observation of exclusive $\chi_c$ and $\gamma\gamma$ events (\cite {MG2,koji,AH})
by the CDF collaboration has been reported at the conferences.
These results appear to be consistent  with the
perturbative QCD expectations \cite {KMRS,KMRSg}, though in reality the scale
of the $\chi_c$ production process is too low to justify the use of the perturbative
QCD formalism.
The Tevatron exclusive $\gamma\gamma$ data are very important.
Here we do not face problems with hadronization or with the identification
of the jets. However the exclusive cross section is rather small.
Future precise measurements in the diphoton mass interval
10-20 GeV would allow a significant reduction of the uncertainties
in the expectations for Higgs production,
to the order of $30-50\%$.}
confirmation of the mechanism of central diffractive production is still desirable.

Here we examine in more detail the prediction for the important process of central diffractive dijet production
at the Tevatron.  This process is a valuable luminosity monitor for central diffractive Higgs production, and for other
exclusive processes which may reveal New Physics, at the LHC.  
The corresponding cross section was evaluated to
be about 10$^4$ times larger than that for the SM Higgs boson.
Thus, in principle, the exclusive production
of a pair of high $E_T$ jets (that is $p\bar {p} \to p+jj+\bar {p}$ in the case of the Tevatron)
appears to be an ideal `standard candle' for the Higgs.
Note, that the CDF measurements have already started to
reach values of the invariant mass of the Pomeron-Pomeron system
in the SM Higgs mass range.
This process is important on its own right
as a gluon factory. As discussed in \cite {KMRmm,KMRProsp}
the remarkable purity of the diffractively
produced di-gluon system
would provide a unique environment to study the properties of high
energy gluon jets.
Unfortunately, in the present CDF experimental environment, which does not provide
tagging of both forward protons, the separation
of exclusive events is not completely unambiguous. In particular, in addition to the smearing  due to the
jet-searching algorithm and
detector effects (see for example, \cite {Royon}) , there are also hadronization and QCD radiative effects,
which distort the manifestation of the exclusive di-jet
signal, see for example \cite {AF, BH75}.
Because the reliability of the predictions
for the cross sections of central exclusive production
of heavy mass objects is so important for the prospects of
forward physics studies at the LHC, it is pivotal to check (whenever possible)
all the important ingredients of the perturbative QCD approach derived
in \cite{KMR, KMRProsp}. In this paper we focus on how to
expose the role of the crucial QCD radiative effects
which regulate the amount of the Sudakov suppression.

Recall, that already in QED, it is well known that we can never observe a pure exclusive process.  For example, the cross section for
$e^-e^+ \to \mu^-\mu^+$ is exactly zero if we exclude the photon
radiation and additional lepton-pair production which may accompany such
events; for a review, see \cite{bfkk}.
To determine the cross section we must use the celebrated Bloch-Nordsiek \cite{BN}
and Kinoshita-Lee-Nauenberg \cite {KLN} theorems, and calculate the radiative correction
accounting for the experimental resolution.  In experiments with very good resolution the corrections are quite large.

An analogous situation occurs when we consider QCD exclusive processes. Here we will apply the Bloch-Nordsieck
procedure to exclusive diffractive dijet production.  That is we will allow for additional gluon radiation
in some rapidity interval $\delta\eta$, and study how the cross section changes as we change the size of $\delta\eta$
and the energy fraction which is allowed to radiate into $\delta\eta$.  At present, two extreme mechanisms are used
to describe central diffractive dijet production.  First, the formalism for pure exclusive production \cite{KMR} has
been implemented in the ExHuMe Monte Carlo \cite{ExHuMe}.  Second, central inelastic dijet production via the
inelastic interaction of two soft Pomerons, which results in parton-parton scattering at large $E_T$; this process
is implemented in the POMWIG Monte Carlo \cite{POMWIG}.  The dijet distribution is plotted in terms of the variable
\be
R_{jj}~=~M_{jj}/M_X~.
\label{eq:jj}
\ee
In terms of this variable, the first process corresponds to $R_{jj}=1$, since the mass of the dijet system, $M_{jj}$,
is equal to the mass, $M_X$, of the whole central system.  The second process has $R_{jj}<1$ since additional
radiation (the fragments of the Pomerons) populate the central region, that is $M_X>M_{jj}$.

\section{A new signature $R_j$ of exclusive dijet events}
Dijet production, with a rapidity gap on either side, has been measured by the CDF collaboration, both in Run I \cite{CDFdijets}
and in Run II \cite {KG2, MG1,CM, MG2, koji}, at the Tevatron.
However there may still be some room for doubt whether {\it exclusive} dijet production,
$p\bar {p} \to p+jj+\bar {p}$, has been actually observed. As mentioned above,
there are various effects which strongly smear the $R_{jj}$ distribution, especially in the absence
of double proton tagging.
The hope was that exclusive events would show up as a peak at $R_{jj}=1$.  Unfortunately the $R_{jj}$ distribution is strongly smeared out by
QCD bremsstrahlung, hadronization, the jet searching algorithm and other experimental effects.  For example, it was shown, using the ExHume Monte Carlo \cite{CP}, that only about $10\%$ of exclusive events with $E_T>7$ GeV have finally $R_{jj}>0.8$, with the CDF cuts used in Run I at the Tevatron.

\begin{figure}
\begin{center}
\includegraphics[height=5cm]{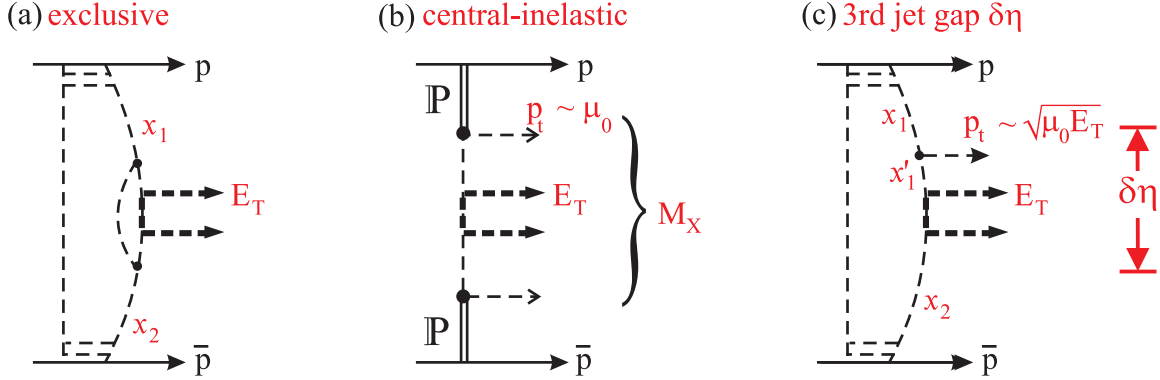}
\caption{Central diffractive dijet production; (a) purely exclusive, (b) via soft Pomeron-Pomeron
interactions, and (c) with a third jet in a given rapidity interval $\delta\eta$.
The dashed lines represent gluons.\label{fig:1}}
\end{center}
\end{figure}

\begin{figure}
\begin{center}
\includegraphics[height=7cm]{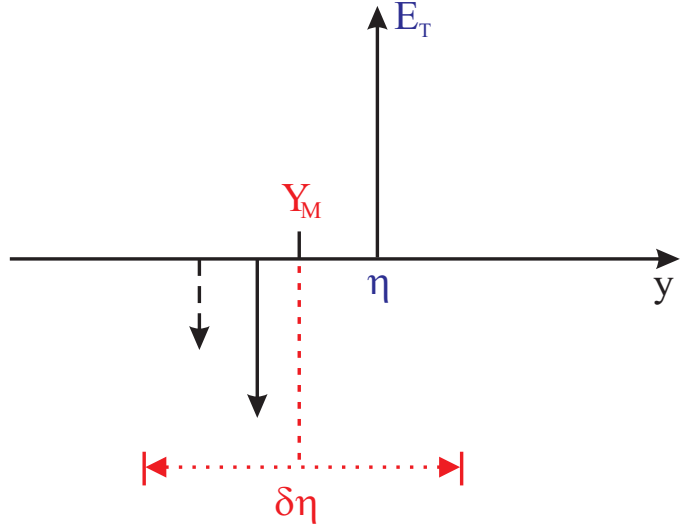}
\caption{The rapidity $Y_M$ of the central system. It does not necessarily occur at $y=0$. The rapidity
interval containing the jets is denoted by $\delta\eta$, outside of which there is no hadronic
activity.\label{fig:2}}
\end{center}
\end{figure}

To weaken the role of this smearing we propose to measure the dijet distribution in terms of a new variable
\be
R_j~=~2E_T ~({\rm cosh}~\eta^*)/M_X~,
\label{eq:j}
\ee
where only the transverse energy $E_T$ and the rapidity $\eta$ of the jet with
the {\it largest} $E_T$ are used in
the numerator.  Here $\eta^* = \eta -Y_M$ where $Y_M$ is the rapidity of the whole central system\footnote{Note
that we systematically neglect the effects arising from the transverse momentum of the dijet system, which is
very small compared to the $E_T$ resolution.}.  Clearly
the jet with the largest $E_T$ is less affected by hadronization, final parton radiation etc.  In particular, final state radiation at the lowest order in $\alpha_S$ will not affect $R_j$ at all, since it does not change the kinematics of the highest $E_T$ jet used to evaluate (\ref{eq:j}). So despite the emission of an extra jet during the final parton shower, we still have $R_j=1$. Thus, to see the role of QCD radiation on the $R_j$ distribution, we only account explicitly for additional gluon radiation
 in the initial state.  At leading order, it is sufficient to consider
the emission of a third gluon jet, as shown in Fig.~\ref{fig:1}.  The reason why it is sufficient to consider
only one extra jet, is that the effect of the other jets, which, at LO, carry lower energy due to the strong ordering,
is almost negligible in terms of the $R_j$ distribution.  The rapidity $Y_M$ is sketched in Fig.~\ref{fig:2}. In Section 5 we will compute the exclusive three-jet cross section for different choices of the rapidity interval $\delta\eta$ containing the jets.

\section{Resume of the calculation of exclusive dijet production}
To compute the $R_j$ distribution we first calculate the cross section of the exclusive dijet production
of Fig.~\ref{fig:1}.  We have $\sigma_{\rm excl}={\cal L}{\hat \sigma}$ where \cite{KMR}
\begin{equation}
{\cal L} ~\simeq ~\frac{{\hat S}^2}{b^2} \left|\frac{\pi}{8} \int\frac{dQ^2_t}{Q^4_t}\: f_g(x_1, x_1', Q_t^2, \mu^2)f_g(x_2,x_2',Q_t^2,\mu^2)~ \right| ^2~.
\label{eq:M}
\end{equation}
The first factor, ${\hat S}^2 $, is the probablity that the rapidity gaps survive against population by secondary hadrons from the underlying event, that is hadrons originating from {\it soft} rescattering. It is calculated using a model which embodies all the main features of soft diffraction \cite{KMRsoft}.  It is found to be ${\hat S}^2 =0.026$ for $pp\ra p+H+p$ at the LHC.
The remaining factor, $|...|^2$, however, may
be calculated using perturbative QCD techniques, since the dominant contribution to the integral comes from the region $\Lambda_{\rm QCD}^2\ll Q_t^2\ll M_H^2$.
The probability amplitudes, $f_g$, to find the appropriate pairs of
$t$-channel gluons ($Q,q_1$) and ($Q,q_2$), are given by the skewed
  unintegrated gluon densities at a {\it hard} scale $\mu \sim M_H/2$.

Since the momentum fraction $x'$ transfered through the
screening gluon $Q$ is much smaller than that ($x$) transfered through
the active gluons $(x'\sim Q_t/\sqrt s\ll x\sim M_H/\sqrt s\ll 1)$, it
is possible to express $f_g(x,x',Q_t^2,\mu^2)$
in terms of the conventional integrated density
$g(x)$. A simplified form of this relation is \cite{KMR}
\begin{equation}
\label{eq:a61}
  f_g (x, x^\prime, Q_t^2, \mu^2) \; = \; R_g \:
\frac{\partial}{\partial \ln Q_t^2}\left [ \sqrt{T_g (Q_t, \mu)} \: xg
  (x, Q_t^2) \right ],
\end{equation}
which holds to 10--20\%
accuracy.
The factor $R_g$ accounts for
the single $\log Q^2$ skewed effect.  It is found to
be about 1.4 at the Tevatron energy and about 1.2 at the energy of the LHC.

Note that the $f_g$'s embody a Sudakov suppression
factor $T$, which ensures that the gluon does not radiate in the
evolution from $Q_t$ up to the hard scale $\mu \sim M_H/2$, and so
preserves the rapidity gaps. The Sudakov factor is \cite{WMR,KMR}
\begin{equation}
\label{eq:a71}
  T_g (Q_t, \mu)=\exp \left (-\int_{Q_t^2}^{\mu^2}
  \frac{\alpha_S (k_t^2)}{2 \pi}\frac{dk_t^2}{k_t^2} \left[
  \int_\Delta^{1-\Delta}zP_{gg} (z)dz
\ + \ \int_0^1 \sum_q\
  P_{qg} (z)dz\right]\right),
\end{equation}
with $\Delta = k_t/(\mu + k_t)$.  The square root arises in
(\ref{eq:a61}) because the (survival) probability not to emit any
additional gluons  is only relevant to
the hard (active) gluon.  It is the presence of this Sudakov factor
which makes the integration in (\ref{eq:M}) infrared stable, and
perturbative QCD applicable.

It should be emphasised that the presence of the double
logarithmic $T$-factors is a purely classical effect, which  was first
discussed in 1956 by Sudakov in QED \cite{Sud}.  There is strong bremsstrahlung
when two colour charged gluons `annihilate' into a heavy neutral object and the
probability not to observe such a bremsstrahlung is given by the
Sudakov form factor.
  Therefore, any model (with perturbative or
non-perturbative gluons) must account for the Sudakov suppression when
producing exclusively a heavy neutral boson via the fusion of two
coloured/charged particles.

In fact, the
$T$-factors can be calculated to {\it single} log
accuracy \cite{KKMRext}. The collinear single logarithms may be summed up using the
DGLAP equation. To account for the `soft' logarithms (corresponding
to the emission of low energy gluons) the one-loop virtual correction
to the $gg\to H$ vertex was calculated explicitly, and then the scale
$\mu=0.62\ M_H$ was chosen in such a way that eq.(\ref{eq:a71})
reproduces the result of this explicit calculation \cite{KKMRext}. It is sufficient to
calculate just the one-loop correction since it is known that the
effect of `soft' gluon emission exponentiates. Thus
(\ref{eq:a71}) gives the $T$-factor to single log accuracy\footnote{ Of course, in the case of QCD, the exponentiation of soft emission requires some clarification.
Because of the non-Abelian structure of QCD, there are indeed some
particular cases when the soft-emission factorization and Poisson
distribution theorems do not hold.  This was exemplified, in
particular, in Ref.~\cite{kf}. However we are
interested in a phenomenon of a completely different (classical) nature.
In \cite{KKMRext} we discussed the NLO correction
 to the double log term caused by the {\it classical current},
where the soft gluon radiation
exponentiates. This accounts for the effect of the
energy- and angular-ordered additional soft gluon radiation, which,
due to QCD coherence, is just part of the cascade generated by
the `primary' gluon.  Summation of such soft `single' logs is
performed analogously to the DGLAP approach, which results in their
exponentiation. This situation is of the same nature as the well
known Modified Leading Logarithmic Approximation, which, for example, is discussed in
detail in the book  by Dokshitzer et al. \cite{DKMT}.}.

\section{Calculation of exclusive 3-jet production}

Here we consider the emission of a third jet described by the variables
$x$ and $p_t$. The variable $x$ is the fraction of the momentum of the incoming gluon (denoted by $x_1$ in Fig.~\ref{fig:1}(c))
carried by the third, relatively soft, jet; that is $x=1-x'_1/x_1$. The explicit formula for the LO third jet radiation can be
obtained using the helicity formalism reviewed in Ref.~\cite{MP}. We outline the calculation
in the Appendix, where the general formulae for the exclusive
three-jet production amplitude are presented; that is, not restricted to LO. In the double logarithm limit, with $p_t \ll E_T$ and
$x \ll 1$, the exclusive 3-jet cross section is simply the exclusive
dijet cross section, ${\hat \sigma}^{(2)}$, multiplied by the classical
probability for soft gluon emission \begin{equation} \label{eq:e1}
d{\hat \sigma}^{(3)}_{\rm LO}~=~d{\hat \sigma}^{(2)}~ \frac{1}{4}~\left(\frac{N_c\alpha_s}{\pi}\frac{dp_t^2}{p_t^2}
\frac{dx}x\right)~.
\end{equation}
Note the extra factor 1/4, which reflects the suppression of soft gluon emission in comparison with the usual classical result given by the expression in brackets.
 Naively we might expect a colour factor $N_c$, but instead we have $N_c/4$. This is due to the absence of the colour correlation between
  the left (amplitude $M$) and the right (amplitude $M^*$) parts of the diagram
  for the cross section, in our case with a colour singlet $s$-channel state.

If we just keep the collinear logs with respect to the beam direction, that is we keep the condition $Q_t<p_t \ll E_T$, but do not impose $x \ll 1$, then the 3-jet cross section becomes
\begin{equation}
\label{eq:e2a}
\frac{d{\hat \sigma}^{(3)}_{LO}}{dt}~=~{\hat \sigma}~
 \left(\frac{N_c\alpha_s}{4\pi}\frac{dp_t^2}{p_t^2}\frac{dx}x\right)\ ,
\end{equation}
where
\be
{\hat \sigma}~=~\left(\frac{9\pi\alpha_s^2(E^2_T)}{4E^4_T}\right)
\frac 12\left[(1-x)^3+\frac{1+x^4(1-2E^2_T/M^2_{jj})}{1-x}\right]\ .
\label{eq:e2}
\ee
The first term, in the round brackets in (\ref{eq:e2}), is the known cross section for the exclusive
 colour-singlet $gg$-dijet production. The variable $t$ in (\ref{eq:e2a}) denotes the square of the four momentum transferred in this exclusive
 colour-singlet $gg\to\mbox{high $E_T$-dijet}$ process. In other words $t$ is measured
 between the highest $E_T$ jet and the incoming gluon which produces the high $E_T$ dijet system.
 The last term in round brackets in (\ref{eq:e2a}) is just the double-log expression for the emission of the third jet, see (\ref{eq:e1}). Finally, the factor in square brackets in (\ref{eq:e2}) 
 accounts for the polarization structure of the 3-jet system. Recall that the exclusive double-diffractive kinematics selects events with the same helicities of the incoming gluons, either ($++$) or $(--)$, that is $J_z=0$. The first term, $(1-x)^3$, corresponds to the helicity of the soft (third) jet being equal to the
 helicities of the incoming gluons, whereas the remaining expression corresponds to the third jet having opposite helicity to that of the incoming gluons. In this expression, the term proportional to $x^4$ originates from the high $E_T$ dijets having different helicities, whereas the
 factor 1 in the numerator corresponds to the production of two high $E_T$ jets with the helicities equal to each other. The $1/(1-x)$ in the second term reflects the usual (BFKL-like) $1/z$ singularity in the Altarelli-Parisi splitting function $P(z)$.

It is informative to note that the behaviour of all three terms in the square
brackets of Eq. (\ref{eq:e2}), in the $x \to 0$ or $x \to 1$ limits, is not accidental.
Its physical origin can be understood by recalling the
celebrated Low soft-bremsstrahlung theorem \cite{Low} (see also \cite{borden,DKS}).
Recall, that according to the MHV rules (see the Appendix),
the only non-vanishing Born $2 \to 2$ amplitudes, $M_B$,
are those which have two positive and two negative helicities.
On the other hand, the $J_z=0$ selection rule requires that the two
incoming gluons have the same helicities, either $(++)$ or $(--)$.
According to the Low theorem \cite{Low}, for radiation of a soft gluon
with energy fraction $z \ll 1$, the radiative matrix element $M_{\rm rad}$
may be expanded in powers of $z$
\be
 M_{\rm rad} \sim  \frac{1}{z} \sum_0^\infty C_n z^n,
\ee
where the first two  terms, with coefficients $C_0$ and $C_1$ (which correspond
to long-distance radiation), can be written
in terms of the non-radiative matrix element $M_B$.

 The application of these classical results is especially transparent when 
the cross sections are integrated over the azimuthal angles. Then the non-radiative
process depends only on simple variables, such as the centre-of-mass energy
\footnote{Note that in our case, in the collinear log approximation, when $Q_t \ll p_t \ll E_T$,
the azimuthal angular dependence is practically absent.}.
In particular, if $M_B =0$, the expansion
starts from the non-universal $C_2 z^2$ term, which corresponds to non-classical
(short-distance) effects, not
related to $M_B$, see \cite{borden,DKS}.

Let us start with the third term in the square brackets in Eq. (\ref{eq:e2}).
In this case soft radiation
should be considered with $z=x \ll 1$.  The corresponding non-radiative matrix element vanishes,
since its helicity structure is either $(+++-)$ or $(---+)$.
Therefore, the matrix element squared, $|M_{\rm rad}|^2$, is proportional
to $x^2$. Keeping in mind the factor $x^2 ~dx/x$, which arises from phase space,
we see that this term is indeed proportional to $x^4~ dx/x$, as it appears
in Eq. (\ref{eq:e2}).
The soft-radiation limit of the first term corresponds to $z=(1-x) \ll 1$.
Then the third jet carries the largest
momentum, and one of the final jets is very soft. Again, the corresponding
Born amplitude vanishes due to the MHV rule,
and we arrive at the result $|M_{\rm rad}|^2 \sim (1-x)^4 ~d(1-x)/(1-x)$.
Finally, the second term, with the factor 1 in the numerator, corresponds to the only non-vanishing
non-radiative amplitude, either $(++--)$ or $(--++)$.

 In the case of the collinear LO process (i.e. $p_t \ll E_T$), the value of $R_j$
 can be calculated as
 \begin{equation}
\label{eq:erj}
R_j=\sqrt{1-x}\left(\frac{\cosh(\eta^*)}{\cosh(\eta^*\pm \frac 12\ln(1-x))}\right)\ .
\end{equation}
Here $\sqrt{1-x}=M_{jj}/M_X$ accounts for a lower mass, $M_{jj}$, of dijet system
in comparison with the mass $M_X$ of 3-jet system, whereas the factor in brackets accounts for the corresponding shift (by $0.5\ln(1-x)$) of the rapidity of dijet system.
The minus sign must be used in (\ref{eq:erj}) when the highest $E_T$ jet goes in the same (beam or target)
hemisphere as the soft (third) jet.

\section{How the third jet affects the distribution in $R_j$}
With knowledge of the luminosity, (\ref{eq:M}), and the cross
  section of the hard subprocess, (\ref{eq:e2}), we can calculate the cross
  section of exclusive 3-jet production, and study how this contribution
  looks in terms of the $R_j$ variable. Note that, after the emission
  of the third jet, the production of other soft jets with $x'<x$
  practically does not alter the value of $R_j$.

\begin{figure}
\begin{center}
\includegraphics[height=5cm]{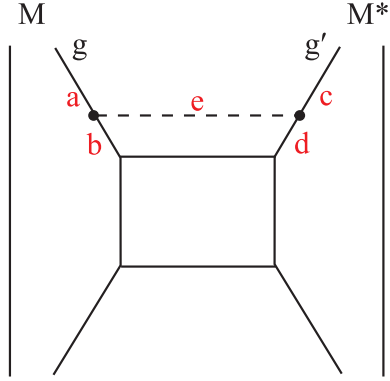}
\caption{The cross section, $MM^*$, for exclusive three-jet production, where the active gluons are denoted by $g$ and $g'$, see the $t$-channel decomposition of eq. (\ref{eq:decomp}). The two outside vertical lines are the screening gluons; indeed all the lines in the plot denote gluons. The dashed line is the third (soft) jet, with kinematic variables $x$ and $p_t$. The colour labels $a,b,c,d,e$ are those used in eq. (\ref{eq:ff}).}
\label{fig:MM}
\end{center}
\end{figure}
In the naive, \`{a} la QED, case, this multijet emission cancels a large part of the Sudakov $T$-factor suppression.
In other words, it gives an exponent analogous to that in (\ref{eq:a71}), but with a
positive power. In QCD the situation is more complicated.  In the
expression for the cross section, $MM^*$, the two active $t$-channel
gluons (one in $M$, the other in $M^*$) are not correlated with each
other, but form colour singlets, each with the corresponding screening
gluon in its own amplitude, $M$ or $M^*$, see Fig.~\ref{fig:MM}. The colour decomposition
of the $t$-channel pair of active gluons, $gg'$, is given by
\be gg'=\sum_i c'_iA_i=\frac {1}{64}A_1+\frac 8{64}A_8+\frac
8{64}A_{\bar 8} +\frac{10}{64}A_{10}+\frac{10}{64}A_{\bar{10}}+\frac
  {27}{64}A_{27}, 
\label{eq:decomp}
\ee
where $A_i$ denotes the colour multiplet of the $t$-channel $gg'$
system: that is, $A_1$ is the colour singlet, $A_8$ and $A_{\bar 8}$ ($A_{10}$ and $A_{\bar{10}}$) are the asymmetric and symmetric colour octets (decuplets) components, etc. The coefficients $c'_i$
give the probability to have one or another colour state. Thus the probability that the pair of active gluons, $gg'$, forms the
corresponding colour multiplet is
\be
c_i \equiv ic'_i,
\ee 
that is $c'_i$ times
the statistical weight given by the number $i$ of members of the multiplet.

If we use the decomposition of the product of two 3-gluon vertices $i^2f_{abe}f_{cde}$ over the colour projection operators $P_i$, that is
\be
i^2f_{abe}f_{cde}=\left(3P_1+\frac 32P_8+\frac 32P_{\bar 8}-P_{27}\right)_{ab,cd} ,
\label{eq:ff}
\ee
then  we see that for each $t$-channel colour multiplet, the probability of soft gluon emission is driven by its own colour factor $\lambda_i$. Namely, we have
  $\lambda_1=N_c=3$ for the singlet, $\lambda_8=3/2$ for the octets,
  $\lambda_{10}=0$ for the decuplets and $\lambda_{27}=-1$ for the 27-multiplet. The colour labels $a,b,c,d,e$ are shown in Fig.~\ref{fig:MM}.

So to compute the $R_j$ distribution we must include the factors arising from including the third jet with the corresponding colour charge $\lambda_i$ for each term in the decomposition (\ref{eq:decomp}). The power of the exponent for this real emission has the form of the $T$ for the virtual corrections (\ref{eq:a71}) multiplied by the corresponding colour factor $\lambda_i/N_c$. For instance, for the case when the $gg'$-pair form a singlet, that is for $i=1$, we have $\lambda_i/N_c=1$. Taking each exponent with its weight $c_i$, we
obtain
\be
T^{(\rm real)}~=~\sum_i c_i~\exp
  \left ( \frac{\lambda_i}{N_c}\int_{Q_t^2}^{\mu^2}
  \frac{\alpha_S (k_t^2)}{2 \pi}\frac{dk_t^2}{k_t^2}
  \int_\Delta^x P_{gg} (z)dz~ \theta(\delta \eta/2-|\eta|)\right)~,
\label{eq:Treal}
\ee
where the scale $\mu=0.62 \sqrt{M^2_{jj}/(1-x)}$ is taken to be the
same as in (\ref{eq:a71}) and where the coefficients $c_i=ic'_i$ are the weightings in the decomposition shown in eq. (\ref{eq:decomp}).
Unlike eq. (\ref{eq:a71}), the $z$ integral is limited by the momentum
fraction $x$ carried by the soft third jet;
for the case of $x>1/2$ the upper limit $x$ in the $z$
integral (14) is replaced by $1-x$ -- two jets cannot carry the fraction
of an initial momentum greater than 1 (i.e. $x+z<1$).
Next, we have added the
  $\theta$-function, which enables us to vary the size of the $\delta\eta$
  interval containing the jets, so that we can study the radiation effect in
  more detail.  As a rule, the jet reconstruction is performed in some
  limited rapidity interval, so it is natural to select events where all
  the jets are emitted within the interval $\delta\eta$ centred at the
  position of the $M_X$ system (that is in the interval $\pm \delta\eta /2$
  in the frame where $Y_M=0$, see Fig.~\ref{fig:2}), while any hadron
  activity outside the interval $\delta\eta$ is forbidden.

  Note that, due to a more complicated colour structure in QCD,
even in the double log limit, there is
  no exact cancellation between the real emission (\ref{eq:Treal}) and
  the Sudakov $T$-factor (\ref{eq:a71})\footnote{The simplest example of this lack of cancellation
  is exclusive Higgs boson production, where already at the
  first $\alpha_s$ order there is Sudakov suppression (\ref{eq:a71}),
  while it is impossible to emit only one gluon accompanying the Higgs
  boson from the colourless two gluon state.}.

\begin{figure}
\begin{center}
\includegraphics[height=15cm]{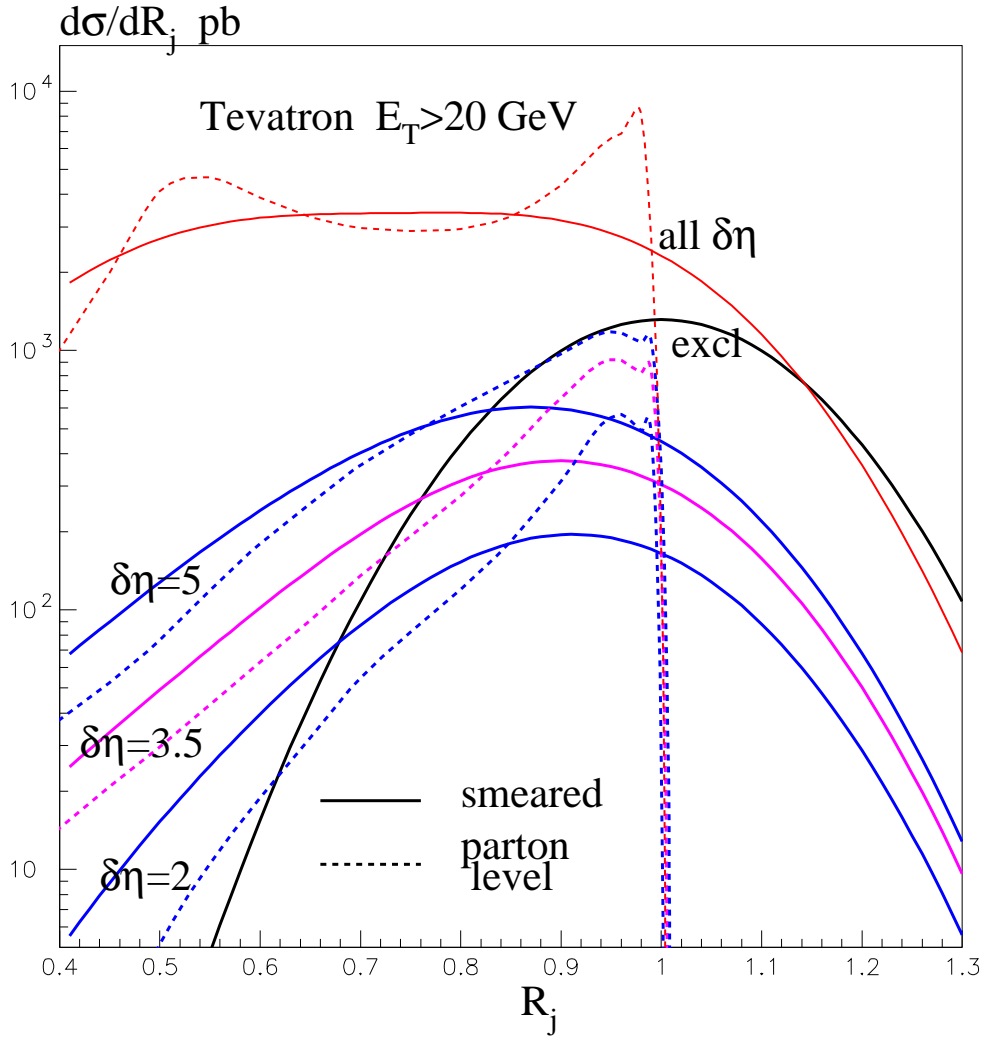}
\caption{The $R_j$ distribution of exclusive two- and three-jet production at the Tevatron.
Without smearing, exclusive two-jet production would be just a $\delta$-function at $R_j=1$.
The distribution for three-jet production is shown for different choices of the rapidity interval, $\delta\eta$, containing the jets; these distributions are shown with and without smearing. The highest $E_T$ jet must have $E_T>20$ GeV.}
\label{fig:teva}
\end{center}
\end{figure}

\begin{figure}
\begin{center}
\includegraphics[height=15cm]{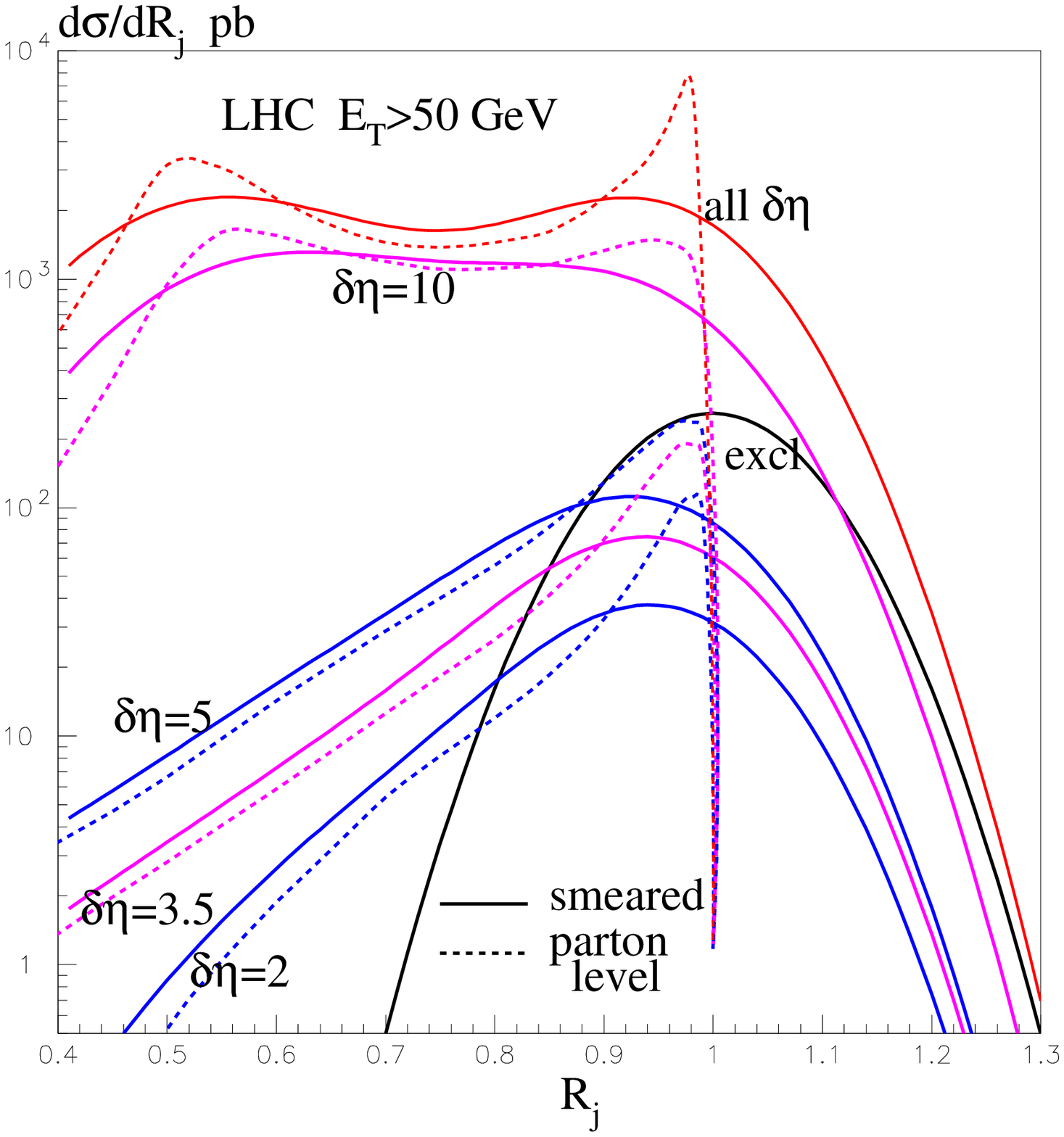}
\caption{The $R_j$ distribution of exclusive two- and three-jet production at the LHC.
Without smearing, exclusive two-jet production would be just a $\delta$-function at $R_j=1$.
The distribution for three-jet production is shown for different choices of the rapidity interval, $\delta\eta$, containing the jets; these distributions are shown with and without smearing. The highest $E_T$ jet must have $E_T>50$ GeV.}
\label{fig:lhc}
\end{center}
\end{figure}

To calculate the exclusive cross section for 3-jet production
accompanied by the emission of softer jets in the rapidity interval
$\delta \eta$, we multiply the exclusive luminosity (\ref{eq:M}) by the cross
section of the hard (LO 3-jet production) subprocess, (\ref{eq:e2}), and by the
factor $T^{({\rm real})}$, (\ref{eq:Treal}), to account for the allowed radiation of
softer gluons. The results are presented in Fig.~\ref{fig:teva} and Fig.~\ref{fig:lhc} in terms of
distributions over the new variable $R_j$.  In order to do this, relation (\ref{eq:erj}) was used to transform the distributions
over the momentum fraction $x$ carried by the soft gluon, into the
$R_j$-distributions presented in the figures.


To be explicit the procedure is as follows. The $R_j$ distribution is computed using
\begin{equation}
\frac{d\sigma}{dR_j}=\int dE_T^2 d\eta_1 d\eta_2 dp^2_t~{\cal L}\left(\frac{d\hat\sigma^{(3)}}{dtdp^2_tdx}
\right)(T^{\rm (real)})^2 \left(\frac{dR_j}{dx}\right)^{-1}
\end{equation}
where the luminosity ${\cal L}$ is given in (\ref{eq:M}) and the Sudakov factor 
$T^{({\rm real})}$ is given by (\ref{eq:Treal}); and where $\eta_1$ and $\eta_2$ 
are the rapidities of the high $E_T$ jets.
We integrate over the kinematic intervals
\be
E_T>E_{\rm min},~~~~~~|\eta_{1,2}|<2.5,~~~~~~ p_{\rm min}<p_t<p_{\rm max}.
\ee
 The lower limit of the logarithmic $p_t$ integral
is given {\it either} by the transverse momentum $Q_t$ in
the gluon loop\footnote{For $p_t < Q_t$, the
destructive interference between emissions from the active gluon $x_1$
and from the screening gluon (that is, the left gluon in Fig.~\ref{fig:1}(a,c)) kills the logarithmic $p_t$
integration. Strictly speaking the values of $Q_t$ in the
amplitudes $M$ and $M^*$ may be different, but this effect is beyond the
LO accuracy of our calculation.}
{\it or} by the allowed rapidity interval $\delta\eta$, that
is $p_{\rm min}={\rm max}\left\{Q_t,xM_Xe^{-\delta\eta/2}\right\}$.
The upper limit is of a pure kinematical nature: $p_{\rm max}={\rm min}\left\{E_T,xM_X/2\right\}$.
If $p_{\rm max}< p_{\rm min}$, then there is no LO contribution.

Next, we have to include the emission of the
third (soft) jet in the direction of one or the other incoming gluons, that is beam
protons. In other words we must sum up the contributions with either the plus or minus signs plus
in (\ref{eq:erj}). Thus, finally, we obtain
$$\frac{d\sigma}{dR_j}=\frac{{\hat S}^2}{b^2}\int dE^2_T
 d\eta_1 d\eta_2\sum_{+,-}\sum_i c_i ~{\hat \sigma}~\left(\frac{N_c \alpha_s}{4\pi}\right)$$
\be
\left|\frac{\pi}{8} \int\frac{dQ^2_t}{Q^4_t}\: f_g(x_1,x_1', Q_t^2,
\mu^2)
f_g(x_2,x_2',Q_t^2,\mu^2)\exp(n_i)\sqrt{\ln (p^2_{\rm max}/p^2_{\rm min})}\right|^2
\left(x\frac{dR_j}{dx}\right)^{-1}
\label{eq:fin}
\end{equation}
where $n_i$ denotes the power in the exponent in $T^{({\rm real})}$ of (\ref{eq:Treal}). The quantity ${\hat \sigma}$ arising from the hard
 $gg\to ggg$ subprocess is given by (\ref{eq:e2}). Note that the factor $dp^2_t/p^2_t$ in (\ref{eq:e2a}) gives rise to the logarithm in ${\cal L}$ in (\ref{eq:fin}), while the factor $dx/x$ goes into $(xdR_j/dx)^{-1}$.  Indeed, the value of $x$ and the derivative
\be
\frac{dR_j}{dx}\ =\ \frac{R_j}{2(1-x)}\left[\pm \tanh\left(\eta^*\pm \frac 12\ln(1-x)\right)\ -\ 1\right]
\ee
are calculated according (\ref{eq:erj}). Note that since
the lower limit, $p_{\rm min}$, of the integration over the $p_t$ of the soft
jet may depend on the transverse momentum, $Q_t$, in the internal gluon
loop, the factors $\exp(n_i)$ and  $\ln(p^2_{\rm max}/p^2_{\rm min})$ occur
inside the `luminosity $Q^2_t$ integral'.

In the computation we have used the partons of Ref. \cite{mrst99}. We neglect
hadronization effects, and present the parton level results by dashed
 curves. In terms of the $R_j$ distribution, the exclusive dijet contribution
 occurs as a $\delta$-function, $\delta(R_j-1)$, and cannot be shown in the
figures. However, in any realistic experiment, the distribution is smeared,
 at least by fluctuations in the calorimeter
\footnote{If we assume that the
two forward protons are tagged,
(as is possible, in principle, in D0 experiment at the
Tevatron\cite{d0,CR} or at the LHC if the CMS and/or ATLAS
detectors are supplemented by the Roman Pots) then the mass of the
whole system, $M_X$, can be measured with much better accuracy by the
missing mass method.}.  To see the effect of more or less realistic
smearing, we assume a Gaussian distribution with a typical
resolution\footnote{We thank M.G. Albrow, D. Alton,
M. Arneodo, A. Brandt, C. Buttar, R. Harris, C. Royon and K. Terashi for discussions on this
choice.
The resolution
$\sigma=0.6/\sqrt{E_T~{\rm in~ GeV}}$
is close to that obtained for the CDF detector, namely
$\sigma=0.64/\sqrt{E_T~{\rm in~ GeV}}+0.028$. The resolution of the D0 hadron
calorimeter is not quite so good: $\sigma\sim 20$\% for $E_T=20$
GeV. Moreover the expected resolution of the CMS hadron calorimeter is about twice worse,
while the anticipated resolution of the ATLAS detector may be even a bit
better: $\sigma\sim 0.5/\sqrt{E_T~{\rm in~ GeV}}+0.015$.} $\sigma=0.6/\sqrt{E_T~{\rm in~ GeV}}$.

  The results obtained,
after this smearing of the parton level distributions, are shown by the
continuous curves in Fig.~\ref{fig:teva} and Fig.~\ref{fig:lhc}. 
We see that for the case of
$\delta\eta < 5$ the  exclusive dijet production still
dominates for $R_j > 0.7\ -\ 0.8$.
The
perturbative QCD radiation is suppressed by the extra coupling
$\alpha_s$. However this suppression is partly compensated by the
collinear logs and by a large longitudinal phase space, that is by the
 rapidity interval $\delta\eta$ allowed for the emission of the extra
soft jets. Indeed, we see that the cross section grows with
 $\delta\eta$, and by $\delta\eta>10$ is close to the saturation curve
 (denoted ``all $\delta\eta$''), which covers the whole interval of
 leading log QCD radiation.

Note that in the region $R_j<0.6-0.7$ the dominant contribution comes from three jet emission. Moreover here the results are more weakly dependent on possible smearing. Of course, in the region of small $R_j$ there may be other contributions coming from the three- or four-jet Mercedes-like configurations\footnote{In the Appendix we give the formulae needed to compute exclusive three-jet production in the whole kinematical interval, and not just in the domain of the leading collinear log approximation.
}.  However these contributions are not expected to be large, since in this case $\alpha_s$ is not compensated by large logs. Another possible contribution comes from configurations which look like inelastic dijet production in the collisions of two soft Pomerons. Such configurations, corresponding to Fig.~\ref{fig:1}(b), may populate the low $R_j$ region, and are beyond the scope of the present analysis.

\section{General use of $R_j$}

In spite of the fact that the $R_j$ variable was introduced to select exclusive
dijets in double-diffractive hadron-hadron interactions in which both of the outgoing protons are tagged, a similar idea can be used to improve the measurements of the light-cone momentum fraction carried by the dijet system in other situations. In particular, to measure the fraction of the photon momentum, $x_\gamma$, carried by the high $E_T$ dijets in  DIS. Note that the final state radiation (and hadronisation)
affect mainly the energy, and much less the rapidity of the jet. Therefore
to calculate $x_\gamma$ (or $x^+_{jj}$ and $x^-_{jj}$ in the more general case)
one can use the $E_T$ of the largest $E_T$ jet together with the rapidity of each jet.

\begin{figure}
\begin{center}
\includegraphics[height=5cm]{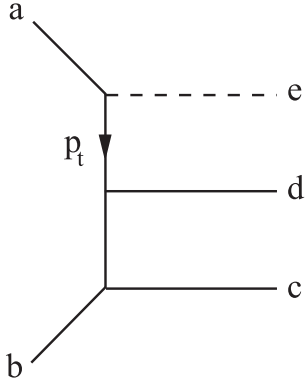}
\caption{A schematic diagram of the $gg \to ggg$ process. The gluon labelled by $e$ is the (soft) third jet.}
\label{fig:5g}
\end{center}
\end{figure}

\section*{Appendix: Helicity amplitudes for $gg \to ggg$}
Here we outline the formalism used to calculate the $gg\to ggg$ process shown in Fig.~\ref{fig:5g}.
We denote the colour indices of the incoming gluons by $a,b$, and of the outgoing high $E_T$ gluons by $c,d$. Finally the colour index of the soft jet is denoted by $e$.
The $gg \to ggg$ matrix element, which depends on the helicities, $h_i$, and the 4-momenta, $p_i$, of gluons, is given by the so-called dual expansion
(see \cite{MP} and references therein)

 \begin{equation}
\label{eq:mhv1}
M^{h_a,h_b,h_c,h_d,h_e}(p_a,p_b,p_c,p_d,p_e)=\sum {\rm Tr} (\lambda_a\lambda_b
\lambda_c\lambda_d\lambda_e)~ m(a,b,c,d,e)\ ,
\end{equation}
where the sum is over the non-cyclic permutations of $a,b,c,d,e$.
The first factor looks as if all the gluons were emitted from the quark loop; 
where $\lambda_i$ are the standard matrices of the fundamental representation of SU(3),
which are normalised as follows
\be
 {\rm Tr}( \lambda^a  \lambda^b)=  \frac{1}{2} \delta^{ab},
\ee
\be
 [\lambda^a,  \lambda^b]= i f_{abc} \lambda^c.
\ee
The colour-ordered subamplitudes, $m(a,b,c,d,e)$, are only functions of the kinematical variables of the process,
i.e. the momenta and the helicities of the gluons. They
may be written in terms of the products of the Dirac bispinors, that is in terms of the angular (and square) brackets
\be
\langle ab \rangle~=~\langle p_a^-|p_b^+\rangle~=~\sqrt{|2p_ap_b|}e^{i\phi_{ab}},
\ee
\be
[ab]~=~\langle p_a^+|p_b^-\rangle~=~\sqrt{|2p_ap_b|}e^{i\bar\phi_{ab}},
\ee
where $2(p_ap_b)=s_{ab}$
is the square of the energy of the corresponding pair.  If both 4-momenta have positive energy, the phase $\phi_{ab}$ is given by
\be
\cos \phi_{ab}=\frac{p_a^xp_b^+-p_b^xp_a^+}{\sqrt{p^+_ap^+_bs_{ab}}},~~~~~~~
\sin \phi_{ab}=\frac{p_a^yp_b^+-p_b^yp_a^+}{\sqrt{p^+_ap^+_bs_{ab}}},
\ee
 with $p^+_i=p_i^0+p_i^z$, while
 the phase $\bar\phi_{ab}$ can be calculated using the identity $s_{ab}=\langle ab \rangle [ab]$.
 Actually the phase $\phi_{ab}$ is irrelevant in our collinear LO calculations,
 except for the fact that $\langle ab\rangle=-\langle ba\rangle$ and
  $[ab]=-[ba]$.  However to calculate the $gg\to ggg$ amplitude beyond
   LO, and to compute a more precise cross section, based on eqs.
   (\ref{eq:mhv1},\ref{eq:mhv2}), we would have to account for the phases.

   Finally, the only non-zero subamplitudes
  \begin{equation}
\label{eq:mhv2}
m(a,b,c,d,e)=ig^3 2^{5/2}\frac{\langle IJ\rangle ^4}{\langle ab\rangle \langle bc\rangle \langle cd\rangle \langle de\rangle \langle ea \rangle}
\end{equation}
are those which have two helicities of one sign, with the other three of the opposite sign,
the so-called Maximal Helicity Violating (MHV) amplitudes. 
Here $g$ is the QCD coupling ($\alpha_s=g^2/4\pi$). In particular, when $h_a=h_b=-1$ while $h_c=h_d=h_e=+1$ the numerator
$\langle IJ \rangle^4=\langle ab\rangle ^4$; i.e. $I$ and $J$ are the only two gluons with the same helicities.
If we change the sign of helicities, then we have simultaneously to replace
the $\langle ij \rangle$ brackets by the $[ij]$ brackets.
Note that the collinear logarithm in the direction of gluon
$a$ comes from the factor $\langle ae \rangle$ (or $\langle
ea\rangle$) in the denominator of (\ref{eq:mhv2}). Thus to obtain the LO
result it is enough to keep only the permutations where the soft gluon
$e$ is close by its nearest neighbour, gluon $a$.

Note that in the formalism leading to (\ref{eq:mhv2}) all the gluons are
considered as incoming particles; that is, the energies of the
gluons $c,d,e$ are negative. In the case when one or two momenta in the product
$\langle ab \rangle$ have negative energy, the phase $\phi_{ab}$ is
calculated with minus the momenta with negative energy, and then
$n\pi/2$ is added to $\phi_{ab}$ where $n$ is the number of negative
momenta in the spinor product.

The three jet cross section (\ref{eq:e2}) is the square of the matrix
element (\ref{eq:mhv1}) calculated using the subamplitudes given by (\ref{eq:mhv2}). In this way, we obtain
\begin{equation}
d\sigma=|M|^2\frac{\delta^{(4)}(\sum_i
p_i)}{64\pi^5s_{ab}}\Pi_j\frac{d^3p_j}{2E_j},
 \end{equation}
where $i=a,b,c,d,e$ and $j=c,d,e$.
To calculate the collinear LO contribution it is enough
 to keep, in (\ref{eq:mhv1}), only the permutations where the soft gluon $e$ is the
nearest neighbour of the incoming gluons $a$ or $b$. For example, for the
case of $e$ collinear to $a$ we need only retain the $m(a,e,b,c,d)$
and $m(a,b,c,d,e)$ subamplitudes, plus the analogous amplitudes
with all the permutations of the gluons $b,c,d$. When we sum over the
permutations of gluons $b,c,d$, and account for the fact that
in collinear approximation the 4-vector $e_\mu$ is parallel to $a_\mu$,
we obtain the exclusive amplitude
of high-$E_T$ dijet production. The factor $\langle ae \rangle$
in the denominator of the subamplitude provides the LO logarithm
$ds_{ae}/s_{ae}$ in the cross section.

\section*{Acknowledgements}

We thank Mike Albrow, Michele Arneodo, Andrew Brandt, Duncan Brown, Brian Cox, Albert De Roeck,
Dino Goulianos, Risto Orava,  Andy Pilkington and
Koji Terashi for useful discussions.
MGR would like to thank the IPPP at the University of Durham for hospitality, and ADM thanks the Leverhulme Trust for an Emeritus Fellowship. This work was supported by the Royal Society,
the UK Particle Physics and Astronomy Research Council, by grants RFBR 04-02-16073, 07-02-00023
and by the Federal Program of the Russian Ministry of Industry, Science and Technology SS-1124.2003.2, and by INTAS grant 05-103-7515.


\end{document}